%% file: Safari.tex
\documentclass[aps,prd,preprint]{revtex4}
\usepackage{bookmark}
\usepackage{hyperref}
\usepackage{graphicx} 

\begin{document}
\title{Cosmological constraints on dark matter particle production rate}
\input{author}
\date{\today}
\begin{abstract}
\input{sec0}

\end{abstract}

\maketitle
\input{sec1}
\input{sec2}
\input{sec5}
\input{sec6}
\input{sec7}
\bibliographystyle{aip}
\bibliography{mybib}
\end{document}

%% file: author.tex
\author{Z. Safari\footnote{z.safari@sci.uok.ac.ir}}
\affiliation{Department of Physics, University of
Kurdistan, Pasdaran St., Sanandaj, Iran}
\author{B. Malekolkalami\footnote{b.malakolkalami@uok.ac.ir}}
\affiliation{Department of Physics, University of
Kurdistan, Pasdaran St., Sanandaj, Iran}

\author{H. Moshafi}
\affiliation{Ibn-Sina Laboratory, Shahid Beheshti 
University, Velenjak, Tehran 19839, Iran}

%% file: sec0.tex
Gravitational particle production has been investigated 
by using Einstein's gravitational field equations in the presence 
of a cosmological constant. To study the mechanism of particle creation, 
the Universe has been considered as a thermodynamics system and 
non-equilibrium thermodynamics has been employed. 
In order to estimate the cosmological parameters with observational 
data, including SNe Ia, BAO, Planck 2015 and HST, we have chosen a 
phenomenological approach for the rate of particle creation.
A non-zero particle production rate was obtained implying 
that the possibility of the particle production is consistent 
with recent cosmological observations.
In the $1 \sigma$ confidence interval, the ratio of $\Gamma/3H_0 $ 
was obtained to be $0.0835  \pm 0.0265$.

%% file: sec1.tex
\section{Introduction}
\label{sec1}
The particle creation mechanism in cosmology was first introduced 
by Schrödinger \cite{Schrödinger}. He investigated the effects of 
particle production on the evolution of the Universe by using  
the microscopic description of the gravitational particle production 
in an expanding Universe. 
About three decades later, Parker and others \cite{Parker, Birrell, Mukhanov, Grib} 
 argued this idea again based on the quantum field theory in curved 
space-time with the motivation to find new consequences of the quantum 
field theory of fundamental particles. Parker combined quantum mechanics 
with general relativity and concluded that the time variations of the 
gravitational field lead to the production of the particle.
After that, Hawking investigated particle production by black holes and 
its compatibility with the laws of thermodynamics \cite{Hawking}.

In most cosmological models, the perfect fluid is taken into account 
while the real fluids are dissipative. Therefore, the description of 
many cosmological phenomena necessitates non-equilibrium or irreversible 
thermodynamics. Eckart \cite{Eckart} and Landau and Lifschitz \cite{Landau1958} 
pioneered the generalization of irreversible thermodynamics from 
Newtonian fluid to relativistic fluid and considered the first order 
deviation from equilibrium. However, the first-order 
theory suffers from stability and causality problems.

The second-order deviation from equilibrium thermodynamics is considered
by \cite{Muller, Israel1976, Pavon, Hiscock}. Altering the dissipative 
phenomena into the dynamical variables that have the causal evolution 
overcome the causality problem, and the evolution equation limits the 
propagation speed of dissipative perturbations.

On the other hand, Prigogine \cite{Prigogine} investigated the open 
thermodynamic system in the context of cosmology and concluded that 
although the particle production has not been achieved from Einstein’s 
gravitational field equations, particle production mechanism is 
consistent with these equations.
In 1992, Calvao \cite{Pavon, Calvao} offered the covariant formulation of 
the particle production mechanisms.

The rate of produced particles should be determined by the quantum field
theory in curved space-time \cite{Birrell}. The exact functional form 
of the particle production rate is still not available; therefore,
cosmologist have adopted the phenomenological approach and fitted it with 
observational data \cite{Steigman, Lima, Lima2014, Ramos}.

In this work, our aim is to constrain the dark matter particle production rate
with the observational data. In Section \ref{sec2}, we apply 
non-equilibrium thermodynamics on the homogeneous and anisotropic 
background and study the particle production 
and its corresponding entropy production.
The cosmological constraints used to estimate the free parameters of 
the model are presented in Section \ref{sec5}.
Eventually, we present the numerical results in Section \ref{sec6}.
Discussion and conclusions are presented in Section \ref{sec7}.

%% file: sec2.tex
\section{Non-equilibrium thermodaynamics and gravitational particle production}
\label{sec2}
Let us consider the expanding Universe with the line element including 
directional scale factors A(t), B(t) and C(t) known as Bianchi type I
 (BI) \cite{Bianchi}  

\begin{equation}
ds^2 = dt^2 - A(t)^2 dx^2 - B(t)^2 dy^2 - C(t)^2 dz^2. 
\label{metric}
\end{equation}
The Einstein gravitational field equations in the presence of the 
cosmological constant are as follows,
\begin{eqnarray}
\frac{\ddot B}{B} + \frac{\ddot C}{C} + \frac{\dot B \dot C}{BC} = \kappa T_{1}^{1} + \Lambda, \label{E1} \\ 
\frac{\ddot A}{A} + \frac{\ddot C}{C} + \frac{\dot A \dot C}{AC} = \kappa T_{2}^{2} + \Lambda, \label{E2} \\
\frac{\ddot A}{A} + \frac{\ddot B}{B} + \frac{\dot A \dot B}{AB} = \kappa T_{3}^{3} + \Lambda, \label{E3} \\
\frac{\dot A \dot B}{AB} + \frac{\dot B \dot C}{BC} + \frac{\dot A \dot C}{AC} = \kappa T_{0}^{0} + \Lambda, \label{E0}
\end{eqnarray}
where $ T^{\mu}_{\nu} $ is the energy-momentum tensor and dot means 
differentiation with respect to the cosmic time t. By introducing 
the time-dependent function $ a(t) $ known as the vacuum scale of the BI Universe 
\begin{equation}
a = \sqrt{-g} = ABC, 
\label{scalefactor}
\end{equation}
one can  write metric functions explicitly. Also by this vacuum scale, 
the generalized Hubble parameter can be obtained
\begin{equation}
\frac{\dot a}{a} = \frac{\dot A}{A} + \frac{\dot B}{B} + \frac{\dot C}{C} := 3H. 
\label{hubble}
\end{equation}
Equations (\ref{E1}-\ref{E3}) with (\ref{hubble}) give
\begin{eqnarray}
\frac{\dot A}{A} - \frac{\dot B}{B}= \frac{X_1}{a}, \nonumber \\ 
\frac{\dot B}{B} - \frac{\dot C}{C}= \frac{X_2}{a}, \nonumber \\ 
\frac{\dot C}{C} - \frac{\dot A}{A}= \frac{X_3}{a}, 
\label{scales}
\end{eqnarray}
where $X_1, X_2, X_3$ are the integration constant. As $ a \rightarrow \infty $,
 there is an isotropic expansion in all directions \cite{Jacobs}. After
the equations (\ref{scales}) are integrated, the explicit expression of the metric functions 
is obtained as follows, \cite{Saha}
\begin{eqnarray}
A(t) = Y_1 ~ a^{1/3} \exp{\left[ X_1 \int{\frac{dt}{a(t)}} \right]}, \nonumber \\ 
B(t) = Y_2 ~ a^{1/3} \exp{\left[ X_2 \int{\frac{dt}{a(t)}} \right]}, \nonumber \\ 
D(t) = Y_3 ~ a^{1/3} \exp{\left[ X_3 \int{\frac{dt}{a(t)}} \right]}.
\end{eqnarray}
Here $Y_i$ and $X_i$ are arbitrary constants that satisfy the following relations:
\begin{equation}
Y_1 Y_2 Y_3 = 1, ~~~~~~~~~~~ X_1+X_2+X_3=0.
\end{equation}
Finally, a little calculation on the equations (\ref{E1}-\ref{E0}) gives the evolution
 equation of $ a(t) $  \cite{Saha}
\begin{equation}
{\dot a}^2 = 3 \left(\kappa  \rho + \Lambda \right) a^2 + C_1.
\label{friedmann}
\end{equation}
Modified energy-momentum tensor for relativistic 
fluid including particle production is given by the following equation
\begin{equation}
T_{\mu \nu} = (\rho +p + \Pi) u_{\mu} u_{\nu} + (p + \Pi) g_{\mu \nu},
\label{emtensor}
\end{equation}
where $u_{\mu}$ is the four-vector of velocity such that $u_{\mu}u^{\mu}=-1$, $\rho$ 
and $p$ are energy density and equilibrium pressure,repectively,  and $\Pi$ is the
pressure associated with the created particle. In a closed thermodynamic system, 
particle number ($ N^{\mu} = n u^{\mu} $) is conserved. Laws of energy conservation 
($T^{\mu \nu}_{~~~; \nu} = 0$) and particle numbers conservation ($N^{\mu}_{~~;\mu} = 0$) 
lead to the following equations
\begin{eqnarray}
\dot n + \theta n = 0, \label{ndot} \\ 
\dot \rho + \theta (\rho +p + \Pi)=0, \label{rhodot}
\end{eqnarray}
where $\theta = u^{\mu}_{~~;\mu}=(\frac{\dot A}{A} + \frac{\dot B}{B} + \frac{\dot C}{C})=3H$, 
and $\dot n = n,_{\mu} u^\mu $. In \cite{Israel} authors introduce the second-order 
non-equilibrium thermodaynamics with entropy flow vector $S^\mu$
\begin{equation}
S^\mu = s N^\mu - \frac{\tau \Pi^2}{2 \zeta T} u^ \mu,
\end{equation}
where entropy per particle is denoted by $s$, fluid temperature by T, 
relaxation time by $\tau$ and coefficient of bulk viscosity by $\zeta$. 
Gibbs equation for this system
\begin{equation}
T ds = d \left( \frac{\rho}{n} \right) + p d \left( \frac{1}{n} \right),
\end{equation}
gives the following relation for variation of the entropy per particle
\begin{equation}
\dot s = - \frac{\Pi \theta}{n T}.
\label{entropyperparticle}
\end{equation}
From the conservation laws (\ref{ndot},\ref{rhodot}) 
and (\ref{entropyperparticle}) it is easy to show that
\begin{equation}
S^{\mu}_{~~; \mu} = - \frac{\Pi}{T} \left[ \theta + \frac{\tau}{\zeta} \dot{\Pi} + \frac{1}{2} \Pi T \left( \frac{\tau}{\zeta T}u^{\mu}  \right)_{; \mu} \right].
\end{equation}
With the following choice for creation pressure
\begin{equation}
\Pi = -\zeta \left[  \theta + \frac{\tau}{\zeta} \dot{\Pi} + \frac{1}{2} \Pi T \left( \frac{\tau}{\zeta T}u^{\mu}  \right)_{; \mu} \right],
\end{equation}
the second law of thermodynamics is satisfied \cite{Subhajit}
\begin{equation}
S^{\mu}_{~~; \mu} =  \frac{\Pi ^2}{\zeta T} \geq 0.
\end{equation}
In open thermodynamic system, the particle number is not preserved 
($N^{\mu}_{~~;\mu} \neq 0$) \cite{Zeldovich, Prigogine2}; thus, 
equation (\ref{ndot}) should be modified as follows,
\begin{equation}
\dot n + \theta n = n \Gamma,
\label{ndot2}
\end{equation}
where $\Gamma$ is the particle production rate. In such a case, the Gibbs equation gives
\begin{equation}
\dot \rho + \theta \left( 1 - \frac{\Gamma}{\theta} \right)(\rho + p) = n T \dot s,
\end{equation}
which means that to recover the energy conservation equation (\ref{rhodot}), 
the entropy per particle must be constant ($ \dot s = 0 $). Gibbs equation with
conservation laws (\ref{rhodot},\ref{ndot2}) gives
\begin{equation}
n T \dot s = -\Pi \theta - \Gamma (\rho + p),
\end{equation}
so under the adiabatic condition, the creation pressure is as follows
\begin{equation}
\Pi = - \frac{\Gamma}{\theta} (\rho + p).
\label{pi}
\end{equation}
Therefore, in the adiabatic process, the creation pressure $\Pi$ 
is linearly related to the particle production rate $\Gamma$.
So under the adiabatic condition, dissipative fluid is equivalent 
to a perfect fluid with variable particle number. On the other hand, 
from equations (\ref{friedmann}, \ref{rhodot}, \ref{pi}) 
it can be deduced that
\begin{equation}
\dot H + \frac{3}{2} H^2(1+\omega) \left(1- \frac{\Gamma}{3H} \right)=0,
\end{equation}
which means that regardless of the amount of equation of state, 
$\Gamma=3H$ leads to a late time de Sitter ($\dot{H}$=0).

In an open thermodynamic system, the entropy change ($dS$) in addition to the 
entropy flow ($d_f S$) involves the entropy production ($d_p S$)  \cite{Harko}
\begin{equation}
dS = d_f S + d_p S, 
\end{equation}
where $d_p S \geq 0$. In the homogeneous Universe, the entropy flow change
is zero and so  the entropy change is only due to the entropy production
\begin{equation}
\frac{dS}{dt} = \frac{d_p S}{dt} = \frac{d(nsV)}{dt} = S \Gamma,
\end{equation}
After integration, we have
\begin{equation}
S(t) = S_0 \exp \left[ 3 \int_{a_0}^{a} \frac{\Gamma}{\theta} \frac{da}{a} \right],
\end{equation}
where $S_0$ and $a_0$ are the present value of entropy and scale factor, respectively.
\begin{figure}
\resizebox{0.5\textwidth}{!}{%
  \includegraphics{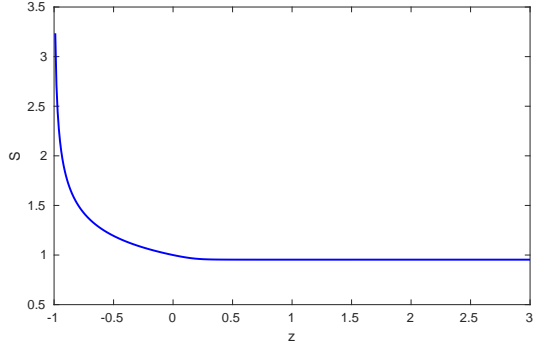}
}
\caption{Entropy changes with $\beta=0.0085$ and $S_0=1$.}
\label{entropy}       
\end{figure}

In the BI Universe dominated by 
pressureless matter, baryonic matter and the energy of the quantum 
vacuum, Friedmann equation (\ref{friedmann})
is as follows
\begin{eqnarray}
\frac{H^2(a)}{{H_0}^2} = \Omega_{b}~ a^{-3} + \Omega_{c}~ a^{-3} \exp \left(3 \int_{1}^{a} \frac{\Gamma}{\theta} \frac{da}{a} \right) \\ \nonumber 
+C_1 a^{-2} + \Omega_{\Lambda},
\label{mymodel}
\end{eqnarray}
only particle production for dark matter is considered here.
To go ahead it is necessary to specify the particle production rate $\Gamma$. 
In fact, $\Gamma$ is determined by studying the irreversible particle
production in quantum field theory in curved space-time. The nature of 
the particles produced affects the particle production rate $\Gamma$ 
since the nature of the dark matter particles is not yet known, a 
phenomenological choice for particle production rate seems to be a 
feasible solution. A general phenomenological choice for particle production rate is 
$\Gamma=3\beta H f(a)$ where $f(a)$ is an arbitrary function of 
the scale factor $a$, and $\beta$ is a non-negative parameter. 
Following Nunes \cite{Nunes}, we work with the phenomenological 
ansatz as follows,
\begin{equation} 
\Gamma = 3 \beta H \left[ 5- 5 \tanh(10- 12 a) \right].
\end{equation}
\begin{figure}
\resizebox{0.5\textwidth}{!}{%
  \includegraphics{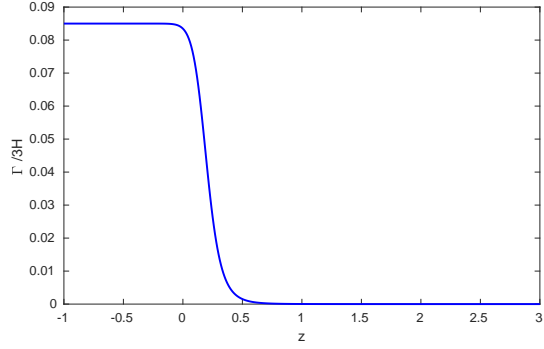}
}
\caption{The ratio of $\Gamma/3H$ in term of redshit with the best fit value of model parameters.}
\label{Gamma3Hz}       
\end{figure}
\begin{figure}
\resizebox{0.5\textwidth}{!}{%
  \includegraphics{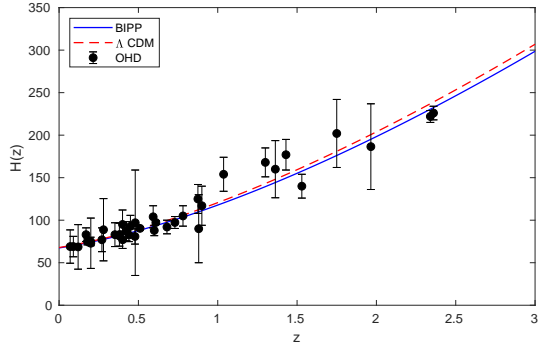}
}
\caption{Theoretical Hubble parameter with the best fit value of model 
parameters along with the observational Hubble data.}
\label{OHD}       
\end{figure}
%

%% file: sec5.tex
\section{Cosmological Constraint}
\label{sec5}
In order to place cosmological constraints on the six free parameters 
of the model (\ref{mymodel}), including 
{\boldmath $\Omega_b, \Omega_c, \Omega_\Lambda, \beta, C_1$} 
and {\boldmath $H_0$}, we run  the CosmoMC package \cite{Lewis2000, Lewis2002}, 
that uses Markov Chain Monte Carlo (MCMC) algorithm to calculate the 
likelihood of cosmological parameters by using SNe Ia, BAO, Planck 2015 
and HST observations. The likelihood function is defined as 
$\mathcal{L} \propto e^{-\chi^2 /2}$, such that $\chi^2$ represents
the difference between observation and theory. The total likelihood 
is obtained by multiplying the separate likelihoods 
of SNe Ia, CMB, BAO, and HST data; thus, 
$\chi^2_{tot} = \chi^2_{SN} +\chi^2_{CMB} + \chi^2_{BAO} + \chi^2_{HST}$. 
For more details about cosmological constraint see \cite{Karami2013, Karami2014}.
\subsection{Type Ia Supernovae (SNe Ia)}
Type Ia supernovae have the same absolute magnitude and therefore 
these standard candles are a powerful tool for exploring the history 
of the expansion of the Universe. For our purpose, we employ 
Joint Light-curve Analysis (JLA) dataset, which in total comprises 
740 SNe Ia data points in the redshift range $0.01\leq z \leq 1.3$ \cite{Betoule}.
118 SNe Ia within the redshift range $0 \leq z \leq 0.1$ from a combination
of various subsamples \cite{Hamuy, Riess, Jha, Contreras, Hicken1, Hicken2}, 
374 SNe Ia from Solon Digital Sky Survey (SDSS) within the redshift range 
$0.3 \leq z \leq 0.4$ \cite{Holtzman}, 239 SNe Ia from the Supernova
Legacy Survey (SNLS) within the redshift range $0.1 \leq z \leq 1.1$ \cite{Guy},
and 9 SNe Ia from Hubble Space Telescope within the redshift range 
$0.8 \leq z \leq 1.3$ \cite{Riess2007} comprise the JLA collection.
\subsection{Baryon Acoustic Oscillations (BAO)}
BAO's standard ruler has provided an other tool to probe the 
expansion history of the Universe. Cosmological perturbations in 
baryon-photon primordial plasma generate pressure waves that affect  anisotropies of the CMB and the large scale structures of matter.
The observed peak in the large scale correlation function measured by 
the luminous red galaxies of Solon Digital Sky Survey (SDSS) at z=0.35 
\cite{Eisenstein} and z=0.278 \cite{Kazin} reveals the baryon acoustic 
oscillations at $100h^{-1}$ Mpc as well as in the two-degree Field 
Galaxy Redshift Survey (2dFGRS) at z=0.2 \cite{Percival}, six-degree 
Field Galaxy Redshift Survey (6dFGRS) at z=0.106 \cite{Beutler}, 
z=0.44, z=0.60 and z=0.73 by WiggleZ team \cite{Blake}, the SDSS 
Data Releases 7 main Galaxy sample at z=0.15 \cite{Ross}, 
the Data Releases 10 and 11 Galaxy samples at z=0.57 \cite{Anderson}. 
\subsection{Cosmic Microwave Background (CMB)}
Acoustic peaks of the temperature power spectrum of the cosmic 
microwave background radiation provide useful information about the 
expansion history of the Universe. The physics of decoupling affects 
the amplitude of the acoustic peaks and the physics of between the present 
and the decoupling changes the locations of peaks. Further on to 
probe the entire expansion history up to the last scattering surface, 
we will include the CMB data from Planck 2015 \cite{Adam, Ade} i.e. a 
joint observation of low\textit{l} + TT temperature fluctuations angular 
power spectrum.
\subsection{Hubble Space Telescope (HST)}
Another independent constraint that can be applied to the estimation of the 
model parameters is the Hubble parameter observational data obtained 
based on different ages of the galaxies \cite{Jimenez}. Because the Hubble 
constant is included in many cosmological and astrophysical calculations, 
NASA/ESA built the Hubble Space Telescope (HST) to measure precise $H_0$, 
and one of the three major HST projects was designated  measurement $H_0$ 
with an accuracy of 10\%.  Freedman and his colleagues have obtained a 
new high-accuracy calibration of the Hubble constant based on our analysis 
of the Spitzer data available to date, combined with data from the Hubble 
Key Project. There was found a value of $H_0 = 74.3$ with a systematic uncertainty 
of $\pm 2.1$ and a statistical uncertainty of $\pm 1.5$ $km ~s^{-1} Mpc^{-1}$ \cite{Freedman2012}.
Riess et al. determined the Hubble constant from optical and infrared 
observations of over 600 Cepheid variables by using the Wide Field Camera 3 
on the Hubble Space Telescope HST. They reported  Hubble constant value of 
$H_0= 73.8 \pm 2.4 km~ s^{-1} Mpc^{-1}$ including systematic errors \cite{Riess2011}. 
 Efstathiou reanalyzed the Riess et al. \cite{Riess2011} Cepheid data using the revised 
geometric maser distance to NGC 4258 of Humphreys et al. \cite{Humphreys}. He concluded 
that $H_0$ based on the NGC 4258 maser distance is $H_0= 70.6 \pm 3.3 km~ s^{-1} Mpc^{-1}$, 
compatible within $1\sigma$ with the recent determination from Planck, also 
assuming that the H-band period-luminosity relation is independent of metallicity 
$H_0= 72.5 \pm 2.5 km~ s^{-1} Mpc^{-1}$ \cite{Efstathiou}. 

%% file: sec6.tex
\section{Numerical results}
\label{sec6}
We have constrained the parameters of the model, using observational data, 
including SNe Ia, BAO, Planck 2015 and HST. In Table \ref{result}, the main 
results of the statistical analysis are summarized.
\begin{table}[!ht]
\caption{ The best fit of the model parameters in $1\sigma$ confidence interval, with Planck TT + 
low\textit{l}, and SNe Ia + BAO + HST + Planck TT + low \textit{l}.}
\centering
\begin{tabular} { l  c  c }
\hline
\hline
 Parameter &~~~~~\begin{tabular}[c]{@{}c@{}}SNe Ia+BAO+HST\\ +Planck TT + low\textit{l}\end{tabular} ~~~~~&  Planck TT+low\textit{l} \\
\hline
{\boldmath$\Omega_b h^2   $} & $0.02192\pm 0.00018        $ & $0.02191\pm 0.00019        $\\

{\boldmath$\Omega_c h^2   $} & $0.11789\pm 0.00096        $& $0.11764\pm 0.00099         $\\

{\boldmath$C_1            $} & $-0.0051\pm 0.0023         $& $-0.0126^{+0.013}_{-0.0086} $\\

{\boldmath$\beta          $} & $0.0085\pm 0.0027          $& $0.0094^{+0.0031}_{-0.0035} $\\

$H_0                       $ & $67.85\pm 0.70             $& $65.3\pm 4.3                $\\

$\Omega_\Lambda            $ & $0.6999\pm 0.0054          $& $0.680^{+0.037}_{-0.028}    $\\
\hline
\hline
\end{tabular}
\label{result}
\end{table}
In the $1 \sigma$ 
confidence interval, the best fit of the $\beta$ parameter is obtained 
$\beta=0.0094^{+0.0031}_{-0.0035} $ for Planck 2015,
$\beta=0.0085 \pm 0.0027$ for joint analysis SNe Ia + BAO + HST + Planck 2015, 
which corresponds to $\Gamma/3H_0 = 0.0923_{-0.0344}^{+0.0304}$
and $\Gamma/3H_0 = 0.0835 \pm 0.0265$, respectively.

This result not only confirms the possibility of particle production, but 
shows that the cosmic scenario involving particle production is consistent 
with observational data. 

Contribution of anisotropy that entered in the 
evolution of the background with parameter $C_1$ is estimated 
$C_1=-0.0126^{+0.013}_{-0.0086}$ for Planck 2015, and
$C_1=-0.0051 \pm 0.0023$ for joint analysis of SNe Ia + BAO + HST + Planck 2015.

The values of Hubble constant $H_0=67.85 \pm 0.7$ for Planck 2015, and
$H_0=67.85 \pm 0.7$ for joint analysis is consistent with Planck 2015 result \cite{Ade}.
The best fit of cosmological parameters, $\Omega_b h^2$, $\Omega_c h^2$, and $\Omega_{\Lambda}$
is obtained $\Omega_b h^2 = 0.02191\pm 0.00019$, 
$\Omega_c h^2 =0.11764\pm 0.00099$, and $\Omega_{\Lambda} = 0.680^{+0.037}_{-0.028}$,
for Planck 2015, and $\Omega_b h^2 = 0.02192 \pm 0.00018$
$\Omega_c h^2 = 0.11789 \pm 0.00096$, and $\Omega_{\Lambda} = 0.6999 \pm 0.0054$
for joint analysis of SNe Ia + BAO + HST + Planck 2015. 
This result is consistent with the values repotred by Planck 2015 \cite{Ade}. 

\begin{figure}
\resizebox{0.5\textwidth}{!}{%
  \includegraphics{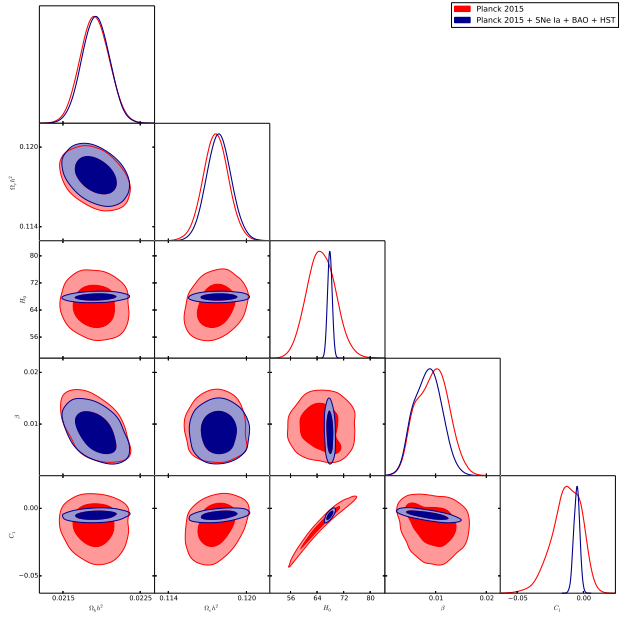}
}
\caption{1D likelihoods for each model parameter, and 2D contours for this
paramaters in $1\sigma$ and $2\sigma$ confidence intervals, with Planck TT + 
low\textit{l} (red), and SNe Ia + BAO + HST + Planck TT + low\textit{l} (blue).}
\label{contour}       
\end{figure}

In Figure \ref{contour}, we have plotted 
the 1D likelihoods and 2D contours for model parameters with 
Planck TT + low\textit{l} (red), and SNe Ia + BAO + HST + Planck TT + low\textit{l}
(blue), where contours represent confidence intervals of 68\% and 95\%.

\begin{figure}
\resizebox{0.5\textwidth}{!}{%
  \includegraphics{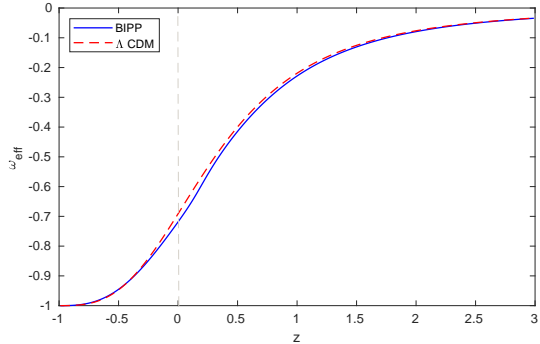}
}
\caption{Effective equation of state with the best fit of 
model parameters along with the effective equation of state of the 
$\Lambda CDM$ model.}
\label{wz}       
\end{figure}

Figure \ref{wz} displays the effective equation of state with the 
best fit of model parameters along with the effective equation of 
state of the $\Lambda CDM$ model. The present value of the effective 
equation of state for this scenario is obtained $\omega_{eff}(z=0)=-0.69$.
As the figure shows, at late time, the effective equation of state for 
this model tends to -1, which corresponds to late time de Sitter. 
In our study, this cosmic scenario showed a deviation 
from $\Lambda CDM$ in the recent past. 

\begin{figure}
\resizebox{0.5\textwidth}{!}{%
  \includegraphics{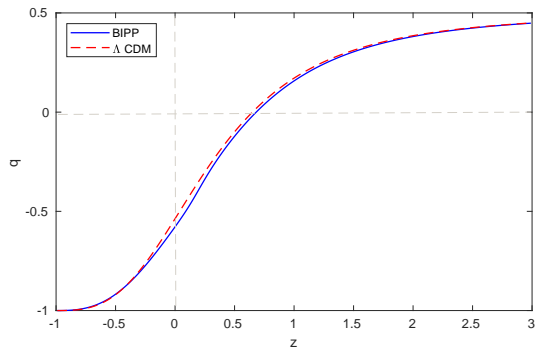}
}
\caption{Deceleration parameter with the best fit of 
model parameters along with the deceleration parameter of the 
$\Lambda CDM$ model.}
\label{qz}       
\end{figure}

Figure \ref{qz} shows the deceleration parameter with the best fit of 
model parameters along with the effective equation of state of the 
$\Lambda CDM$ model. The present value of deceleration parameter 
for this scenario is obtained $q(z=0)=-0.65$. The transition from 
deceleration to acceleration era occurred in $z_t=0.69$. 
Our analysis of this cosmological model revealed a deviation from 
the $\Lambda CDM$ model in the recent past. 

\begin{figure}
\resizebox{0.5\textwidth}{!}{%
  \includegraphics{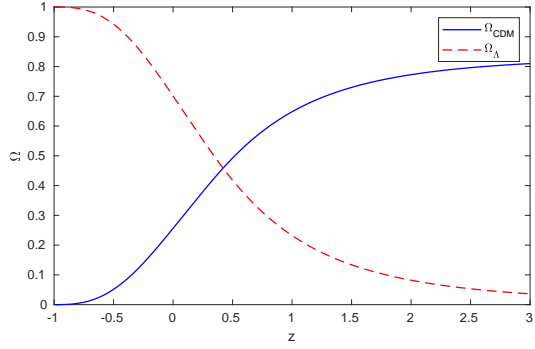}
}
\caption{he best fits of the dimensionless density parameters
of CDM, $\Omega_{CDM} = 8\pi G\rho_{CDM}/3H^2$ $\Lambda$, 
$\Omega_{\Lambda} = 8\pi G\rho_{\Lambda}/3H^2$
using the full data sets.}
\label{Omega}       
\end{figure}

In Figure \ref{Omega}, we plotted the evolutionary behaviors of 
the density parameters of Cold Dark Matter, 
$\Omega_{CDM} = \frac{8\pi G\rho_{CDM}}{3H^2}$, and vacuum density,
$\Omega_{\Lambda} = \frac{8\pi G\rho_{\Lambda}}{3H^2}$. The figure 
displays the fact that as the density parameter of CDM is decreased, 
the vacuum density is increased, during the history of the Universe.
%

%% file: sec7.tex
\section{Coclusions}
\label{sec7}
In this paper, we examined the particle creation from the perspective of non-equilibrium 
thermodynamics in Bianchi type I Universe and tried to explain the evolution of 
the Universe by choosing a phenomenological approach for the particle production 
rate. 

To constrain the free parameters of the model, we use the CosmoMC package. 
Our analysis includes observational data such as supernovae type Ia from JLA, 
cosmic microwave background from Planck 2015, baryon acoustic oscillation from SDSS, 
2dFGRS, 6dFGRS, and observational Hubble data from HST.
The results are as follows:

\begin{itemize}
\item[$\ast$] Figure \ref{entropy} shows that entropy production has started to increase 
near the present time. Figure \ref{Gamma3Hz} shows that in early times 
$\Gamma/3H \ll 1$ and such as the entropy production near the present 
starting to increase and at late time $\Gamma/3H < 1$.

\item[$\ast$] In $1\sigma$ confidence interval the best fit of the $\beta$ parameter is 
obtained $\beta = 0.00 \pm 0.00$ for Planck 2015,
$\beta = 0.0085 \pm 0.0027$ for Planck 2015 + SNe Ia + BAO + HST. This value 
of the $\beta$ implies that $\Gamma/3H = 0.0 $ and $\Gamma/3H = 0.0835$, respectively. 
Thus, the possibility of particle production is approved as consistent with 
recent cosmological observations.

\item[$\ast$] Contribution of anisotropy in this model entered with $C_1$ obtained 
for Planck 2015 data is $C_1=-0.00 \pm 0.00$ and for joint analysis 
Planck 2015 + SNe Ia + BAO + HST data  is $C_1=-0.0051 \pm 0.0023$.

\item[$\ast$] In this cosmic scenario the best fit of the cosmological parameters, 
$\Omega_b h^2 = 0.02192\pm 0.00018$, $\Omega_c h^2 = 0.11789\pm 0.00096$, 
and $\Omega_\Lambda = 0.6999\pm 0.0054$ is consistent with similar values 
in the $\Lambda CDM$ model. Figure \ref{contour} displayed 1D likelihoods 
and 2D contours for model parameters with Planck TT + low \textit{l} (red), 
and joint analysis SNe Ia + BAO + HST + Planck TT + low \textit{l} (blue), 
in $1\sigma$ and $2\sigma$ confidence intervals.

\item[$\ast$] Effective equation of state for Planck 2015 data is 
$\omega_{eff}(z=0)=-0.0$, and for joint analysis Planck 2015 + SNe Ia + BAO + HST 
data is $\omega_{eff}(z=0)=-0.69$. Figure \ref{wz} shows EoS changes it is clear 
that at late time, it behaves like
the $\Lambda CDM$ model. 

\item[$\ast$] The deceleration parameter q reveals a transition from an early matter-dominant 
($q=0.5$) epoch to the de Sitter era ($q=-1$) at late time, as expected. 
The present value of the accelerating epoch starts at transition redshift 
$z_t=0.69$. The deceleration parameter $q(z=0)=-0.65$ obtained.

\item[$\ast$] Evolutionary behaviors of the density parameters of CDM, 
$\Omega_{CDM} = \frac{8\pi G\rho_{CDM}}{3H^2}$, and vacuum density,
$\Omega_{\Lambda} = \frac{8\pi G\rho_{\Lambda}}{3H^2}$ plotted in Figure \ref{Omega},
show that as the $\Omega_{\Lambda}$ is increased, $\Omega_{CDM}$ is decreased, 
during the history of the Universe.
\end{itemize}